**В. І. Марсакова[1,2], І. Л. Андронов[3], В. О. Борщенко[3],**
**І. А. Гарбажій-Романченко[1], А. Д. Лашкова[4],**
**С. А. Кремінська[3], П. А. Дубовски[5],**
**В. В. Дубовський[6], К. Петрік[7]**

[1]Рішельєвський науковий ліцей, Одеса
вул. Університетська, 5, Одеса, Україна, 65026
[2]Головна астрономічна обсерваторія Національної академії наук України,
вул. Академіка Заболотного, 27, Київ, Україна, 03143
[3]Одеський національний морський університет,
вул. Мечникова, 34, Одеса, Україна, 65029
[4]Університет Павола Йозефа Шафарика,
вул. Штробарова, 2, 04180, Кошице, Словаччина
[5]Вігорлатська обсерваторія у Гуменному,
вул. Мієрова, 4, 06601, Гуменне, Словаччина
[6]Одеський національний університет ім. І. І. Мечникова,
вул. Всеволода Змієнка, 2, 65000, Одеса, Україна
7Обсерваторія та планетарій ім. М.Р. Стефаника,
вул Сладковіцова, 41, Глоговець, Словаччина
E-mails: vlada.marsakova@gmail.com, tt_ari@ukr.net, vict0riaborschenko@gmail.com,
ilya.garbazhiy@gmail.com, var@kozmos.sk

# Визначення параметрів кривих блиску малодосліджених затемнюваних змінних зір з використанням даних TESS та інших оглядів неба

*Досліджувалась група маловивчених затемнюваних змінних (з невпевненою класифікацією та/або з невідомими періодами змін блиску) з використанням фотометричних спостережень космічної місії TESS та даних з оглядів неба NSVS, ASAS-SN. Виконано також спостереження мінімумів блиску деяких зір на телескопах Астрономічної обсерваторії у Гуменному на Колоницькому сідлі (Словаччина) та Обсерваторії та планетарію у Глоговці (Словаччина) під час астрономічного табору «Variable-2024». Було уточнено періоди змінності зір, визначено їхні параметри та класифікацію. Для змінних NSV 575 та NSV 014 періоди було знайдено вперше, але залишається сумнів, що NSV 014 — затемнювана змінна, оскільки у неї не було знайдено за-*

*темнень, а спостерігалась тільки асиметрична хвиля, тобто ця змінна зоря може бути перекласифікована як малоамплітудна пульсуюча. Для апроксимації кривих блиску та визначення таких параметрів зоряних систем як глибини та тривалості затемнень, величини ефектів відбиття та еліптичності зір були використані різні методи. Для первинної оцінки періодів використовувався метод періодограмного аналізу, заснований на апроксимації тригонометричним поліномом високих степенів (до 10-го). Для кращої апроксимації повної затемнюваної фазової кривої блиску використовувався алгоритм «New Algol Variable». Методи «Асимптотичних парабол» та «Асимптотичних парабол з підтримувальними стінами» були використані для обчислення моментів затемнень, вони апроксимують не повну кривy, а лише частину кривої поблизу затемнення. Ці методи було впроваджено поряд з іншими у програмі MAVKA. Для змінних NSV 489 та NSV 1884 побудовано діаграми O – C з використанням наших моментів затемнень та моментів, знайдених у літературі. Для NSV 489 період був відкоригований з урахуванням нахилу діаграми O – C.*

## ВСТУП

У каталозі VSX (International Variable Star Index [15]), NSV 575 та NSV 014 позначені як EA: (непевна класифікація як змінні типу Алголя) без значень періодів. Їхня змінність була запідозрена Гофмейстером [13, 14] з амплітудою змін яскравості у фотографічному діапазоні 14.5… $15.0^m$ та $11.5…12.5^m$ відповідно. Дані TESS підтверджують, що NSV 575 — затемнювана змінна, тоді як NSV 014 показує плавні асиметричні зміни блиску. Для них обох ми вперше знайшли періоди змінності.

Зоря NSV 489 була відкрита Ріхтером [19], класифікована як EA [10] з вказаним у VSX періодом $2.8266^d$ та епохою HJD 2455924.30705. Вона також ідентифікована як змінна у GAIA DR3 [12].

Зорю NSV 1884 відкрив Гофмейстер [14], а у роботі [11] вона вивчалась за даними Catalina Real-Time Transient Survey. У каталозі VSX для неї приведено період $1.770432^d$ та початкову епоху HJD 2455230.0.

## СПОСТЕРЕЖЕННЯ

Ми спиралися на використання спостережень TESS [20] для кожної змінної, оскільки вони мають відмінну точність та часове розділення. Для NSV 575 ми знайшли також деякі спостереження ASAS-SN [16], а для NSV 489 та NSV 1884 ми проаналізували дані з архіву NSVS [23]. NSV 489 спостерігалася також під час астрономічного табору «Varia-

ble-2024», організованого Астрономічною обсерваторією на Колоницькому Сідлі (Словаччина), за допомогою 60-см телескопа (ZC600) Обсерваторії та Планетарію у Глоговці (з віддаленим доступом). Наші нові спостереження у Словаччині були виконані у спектральній смузі V поблизу моментів затемнень.

## МЕТОДИ АНАЛІЗУ

Щоб знайти період коливань блиску, ми використовували періодограмний аналіз, заснований на апроксимації тригонометричним поліномом степеня $s$:

$$m(t) \quad C_1 \quad {}_{j\;\;1}^{s}(C_{2j}\cos jwt \quad C_{2j\;\;1}\sin jwt).$$

Тут $w = 2\ \ f$ — кутова частота, $f = 1/P$ — частота, $P$ — період. Незважаючи на поширену назву «періодограмний аналіз», його типовим представленням є залежність тест-функції від частоти. Початковий алгоритм та програма були описані Андроновим [1] та реалізовані у програмі MCV [6].

Для пульсуючих зір попереднє значення дорівнює $s = 1$, а для затемнюваних змінних рекомендована величина є значно більшою та залежить від ширини $D$ затемнення в одиницях періоду, тобто звичайно вона пропорціональна $1/D$. Для затемнюваних змінних розподіл яскравості є асиметричним (спостереження лише зрідка потрапляють на затемнення, натомість більшість даних отримано поза затемненнями), таким чином значення $s$ має бути більшим (попередньо 6…10, з наступним визначенням статистично оптимальної величини). Рекомендований крок за пробними частотами дорівнює $f \qquad / (s \quad T)$, де $\quad < 0.1$, $\quad T = \max(t) - \min(t)$. Для малих кількість пробних частот зростає разом з часом обчислення, роблячи розподіл точок на періодограмі щільнішим. Для аналізу Фур'є $\quad = 1 - 1/n$, де $n$ — кількість точок. Після вибору на періодограмі частоти, яка відповідає найбільшій величині тест-функції, виконується наступний крок: робиться апроксимація для різних $s$ та визначається «статистично оптимальна» величина згідно з критерієм Фішера. Для кожної величини $s$ застосовується нелінійна апроксимація, за допомогою якої також визначається період $P$ [1].

Отже, цей метод був застосований для попереднього визначення періоду змінності вивчених зір. Оскільки NSV 014 показує циклічну хвилю, але не показує змінності, типової для затемнень, ми використовували апроксимацію тригонометричним поліномом для фазової кривої їхнього блиску.

Для затемнюваних змінних з вузькими мінімумами статистично оптимальні величини можуть сягати досить великих значень (десятки), що може призводити до несправжніх хвиль на кривій блиску. Разом з тим внаслідок використання плавних функцій (синуси та коси-нуси), неможливо визначити ширину затемнення (яка є важливим параметром для каталогів), оскільки затемнення має певний початок та кінець.

Щоб зменшити кількість параметрів, які потрібно визначити, І. Л. Андронов [3] запропонував метод та програму «New Algol Variable (NAV)», яку ми використовували для зір з даної вибірки. Подальший аналіз для інших зір показав, що ця апроксимація добре застосовується не тільки для змінних типу Алголя (EA), але й для зір типів Lyr (EB) та W UMa (EW) [21]. Основна ідея полягає в тому, щоб замінити тригонометричні поліноми степеня $s$ сумою позазатемнюваного «континууму» ($s = 2$) та локальних функцій, які описують різні профілі первинного та вторинного мінімумів. Варіації цього алгоритму були вивчені Мікулашеком [18] та І. Л. Андроновим із колегами [7]. А втім, додаткові параметри не поліпшують цю апроксимацію суттєво, тому ми використовували первісний алгоритм NAV.

Для визначення індивідуальних моментів мінімумів (за спостереженнями TESS та нашими) ми використовували програму MAVKA [9], яка доступна за адресою: http://uavso.org.ua/mavka. Вона дозволяє вибирати найбільш точний момент мінімуму з апроксимацій 22 функціями 11 типів. Для обчислення моментів затемнень досліджуваних затемнюваних змінних (NSV 575, NSV 489 та NSV 1884) ми вибрали методи асимптотичних парабол (Asymptotic Parabola [17]) та асимптотичних парабол з підтримувальними стінами (Wall Supported Asymptotic Parabola [8]) через найкращу точність апроксимацій майже V-образних мінімумів.

Ці та інші методи були описані І. Л. Андроновим у роботах [2, 4].

## ПАРАМЕТРИ NAV

Параметри, визначені за допомогою алгоритму NAV, представлено у табл. 1, а відповідні фазові криві приведено на рис. 1. Як можна побачити, алгоритм NAV, застосований до спостережень TESS, дозволив виміряти для деяких з цих змінних типу Алголя тонкі ефекти, такі як ефект відбиття (NSV 575 та NSV 489) та ефект еліптичності зір (NSV 489 та NSV 1884). У таблиці збережено значення параметрів, які не перевищують три стандартні відхилення (виділено жирним шрифтом). Для NSV 489 вторинне затемнення є суттєвим тільки за даними TESS. Детальний опис всіх параметрів, які можна розрахувати за допомогою алгоритму NAV, представлено у роботі [21].

Стандартне відхилення періодів більше для даних TESS, тому що повний інтервал часу цих спостережень є меншим (хоча часове розділення вище), ніж відповідні інтервали для даних NSVS та ASAS-SN.

***Таблиця 1.*** **Параметри затемнюваних кривих блиску, обчислені за допомогою алгоритму NAV**

| Параметр | NSV 575 | | NSV 489 | | NSV 1884 | |
|---|---|---|---|---|---|---|
| | ASAS-SN | TESS | NSVS | TESS | NSVS | TESS |
| Період з VSX | — | — | 2.8266 | 2.8266 | 1.770432 | 1.770432 |
| Період NAV | 0.9535787 ±0.0000012 | 0.9535877 ±0.0000028 | 2.826627 ±0.000004 | 2.8266719 ±0.0000084 | 1.7704263 ±0.0000044 | 1.770317 ±0.000027 |
| Початкова епоха 2400000+ | 57570.45176 ±0.00061 | 58338.083385 ±0.000081 | 57753.1455 ±0.0015 | 58968.611969 ±0.000066 | 57499.6689 ±0.0023 | 58450.39089 ±0.00022 |
| Глибина первинного мінімуму | 0.393 ±0.014 | 0.3348 ±0.0019 | 0.3676 ±0.0105 | 0.6492 ±0.0013 | 0.376 ±0.022 | 0.6338 ±0.0042 |
| Глибина вторинного мінімуму | 0.118 ±0.015 | 0.1455 ±0.0018 | **0.0273 ±0.0211** | 0.0713 ±0.0010 | 0.277 ±0.024 | 0.4998 ±0.0044 |
| Півширина затемнення | 0.0542 ±0.0021 | 0.05509 ±0.00031 | 0.0457 ±0.0019 | 0.039771 ±0.000085 | 0.0717 ±0.0032 | 0.07732 ±0.00045 |
| Ефект відбиття | **0.0094 ±0.0028** | 0.00348 ±0.00034 | **0.0070 ±0.0029** | 0.01086 ±0.00019 | **–0.0080 ±0.0060** | **–0.0067 ±0.0010** |
| Ефект еліптичності | **–0.0021 ±0.0029** | **0.00092 ±0.00037** | **0.0026 ±0.0030** | 0.00190 ±0.00019 | **0.0141 ±0.0071** | 0.0245 ±0.0012 |
| Кількість точок даних | 540 | 1061 | 164 | 1012 | 196 | 1170 |

## NSV 14 — НЕ ЗАТЕМНЮВАНА?

NSV 14, ймовірно, не є затемнюваною, бо ми не знайшли затемнень в даних TESS. Ми використали апроксимацію тригонометричним поліномом другого ступеня та періодограмний аналіз для пошуку періоду та початкових епох і отримали такі значення:

$$P = 13.1 \quad 0.2^d$$

$$T_{0\min} = 58365.3 \quad 0.1 \text{ (для мінімумів)}$$

$$T_{0\max} = 58371.8 \quad 0.1 \text{ (для максимумів).}$$

NSV 14 показує плавну асиметричну хвилю малої амплітуди ( $0.002^m$, див. рис. 2), таким чином, вона може бути пульсуючою змінною.

## МОМЕНТИ МІНІМУМІВ ТА ДІАГРАМИ *O – C*

*NSV 575.* Ми обчислили два середні моменти мінімумів, використовуючи дані TESS та ASAS-SN та застосовуючи до них алгоритм NAV (табл. 2). Є сподівання, що вони будуть корисними для подальшого аналізу стабільності періоду цієї змінної. Ми також можемо викорис-

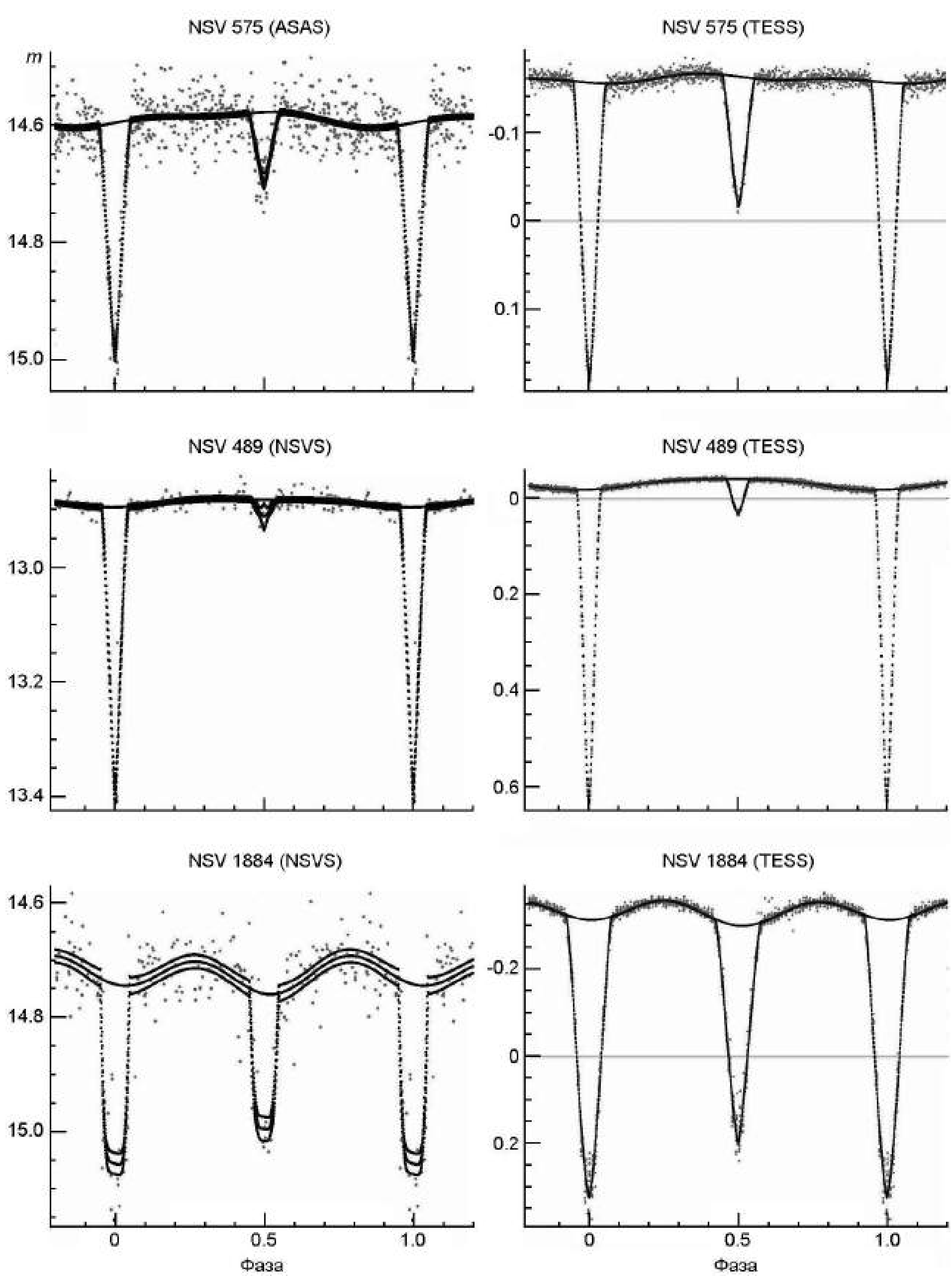

*Рис 1*. Фазові криві блиску, апроксимовані методом NAV

тати ці два моменти для обчислення усередненого значення періоду. Оскільки між ними 805.00081 орбітальних циклів, то усереднене значення періоду дорівнює 0.9535797  0.00000076$^d$.

*NSV 489 та NSV 1884*. Мінімуми за даними NSVS були отримані як середні за апроксимацією алгоритмом NAV. За даними TESS та спостереженнями на телескопі ZC600 ми розрахували індивідуальні

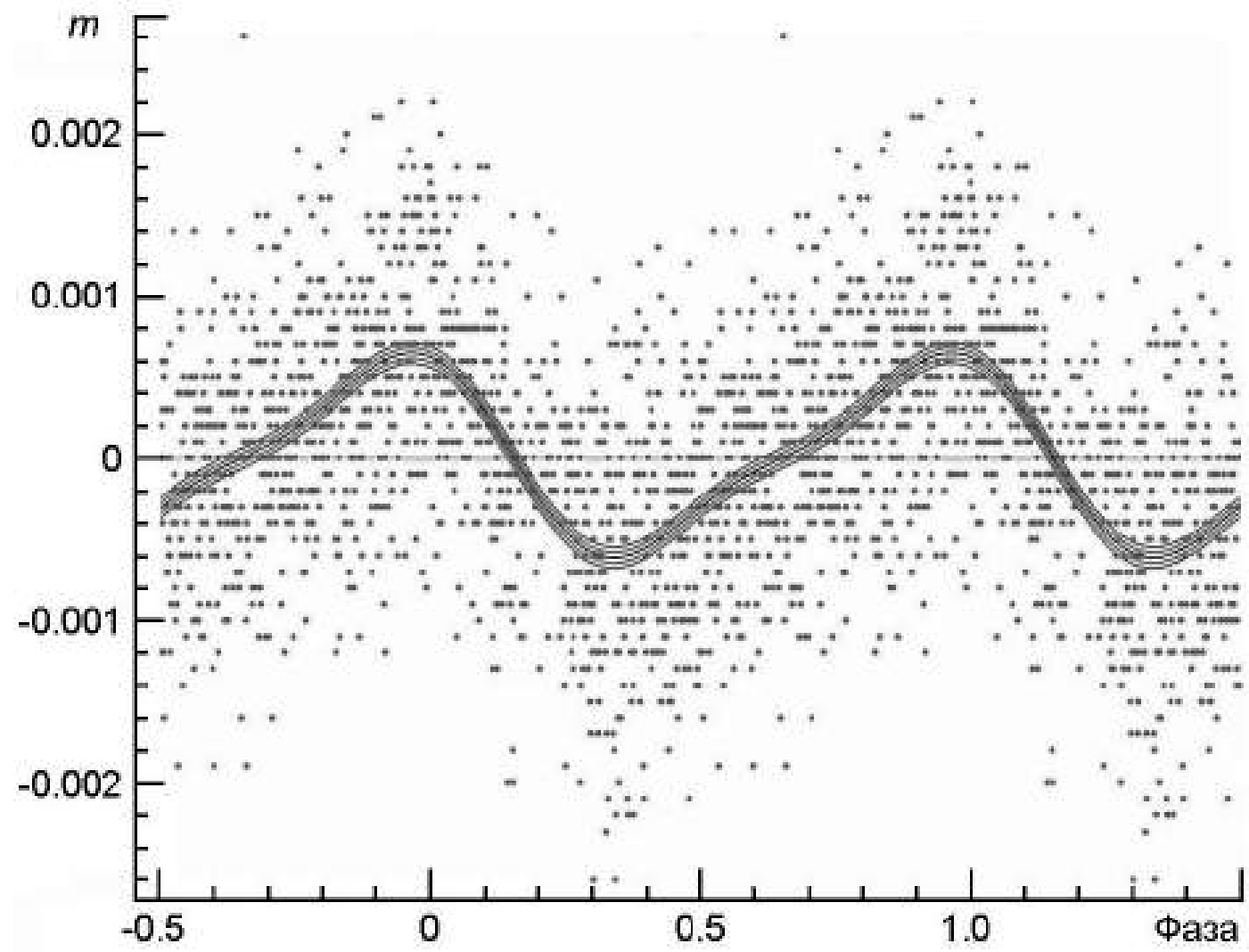

*Рис 2.* Фазова крива блиску NSV 14 та її апроксимація тригонометричним поліномом другого ступеня

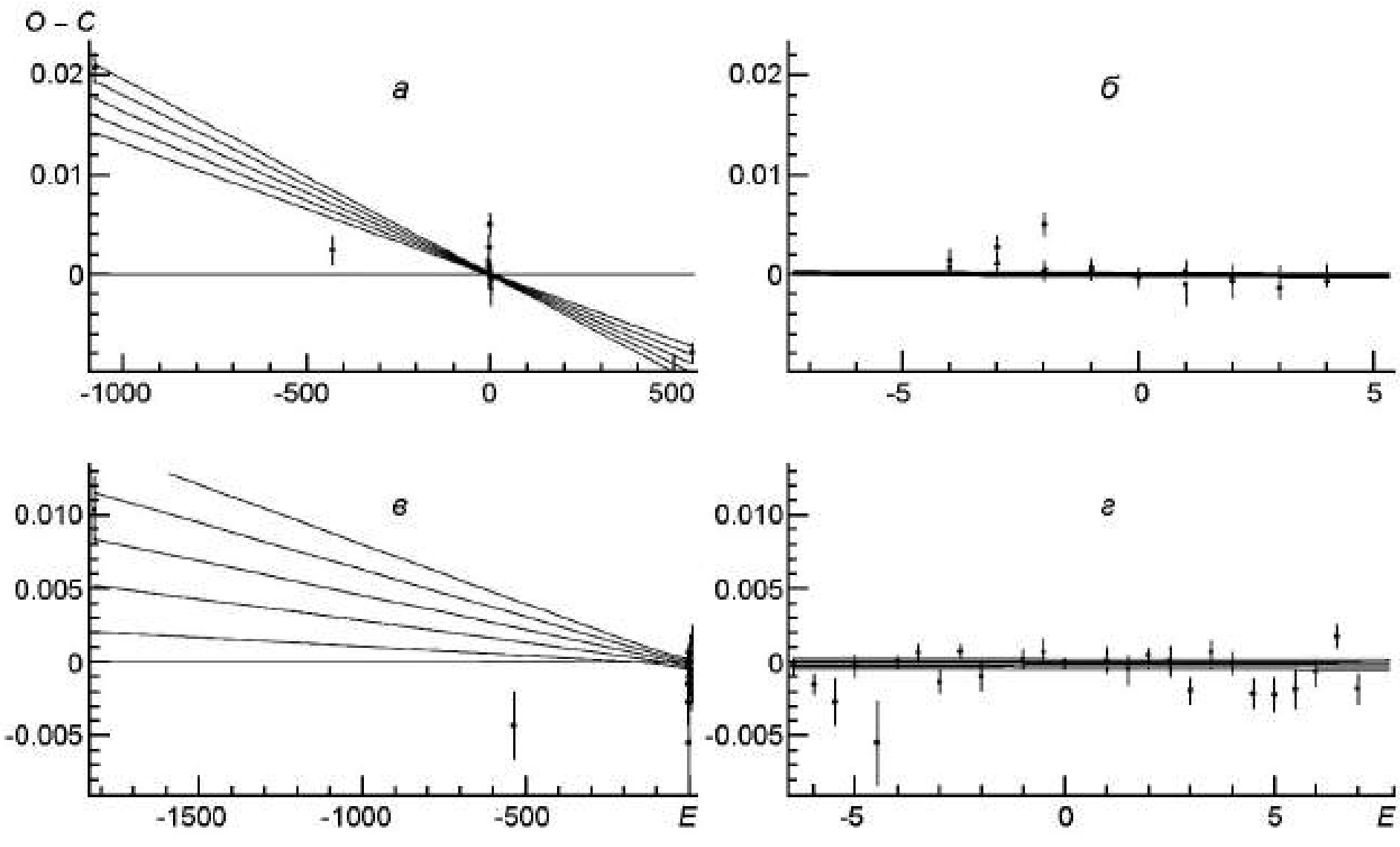

*Рис. 3.* Діаграми *O – C* для NSV 489 (*а, б*) та NSV 1884 (*в, г*): всі мінімуми — *а, в,* мінімуми TESS — *б, г*

моменти мінімумів для кожного затемнення, за допомогою програми MAVKA та використовуючи методи асимптотичних парабол (AP) [17] для первинних та вторинних мінімумів NSV 489 та асимптотичних парабол з підтримувальними стінами (WSAP) [8] для мінімумів NSV 1884.

Оскільки для моментів мінімумів, опублікованих у VSX, немає оцінок похибок, ми взяли для них наші типові похибки.

Діаграма *O – C* для NSV 489 (рис. 3*а*, 3*б*) показує невеликий лінійний тренд, отже можлива мінімальна корекція значення періоду на

***Таблиця 2.*** **Моменти мінімумів**

| HJD | | Min I/II | джерело | HJD | | Min I/II | джерело |
|---|---|---|---|---|---|---|---|
| | NSV 575 | | | | NSV 1884 | | |
| 2457570.45176 | 0.00061 | I | ASAS-SN | 2455230.00 | | I | VSX |
| 2458338.08339 | 0.00008 | I | TESS | 2457499.6689 | 0.0023 | I | NSVS |
| | | | | 2458450.39089 | 0.00022 | I | TESS (NAV) |
| | NSV 489 | | | 2458438.8829 | 0.0006 | II | TESS |
| 2455924.30705 | | I | VSX | 2458439.7669 | 0.0007 | I | TESS |
| 2457753.1455 | 0.0015 | I | NSVS | 2458440.6509 | 0.0016 | II | TESS |
| 2458968.611969 | 0.000066 | I | TESS (NAV) | 2458441.5385 | 0.0008 | I | TESS |
| 2458957.3061 | 0.0009 | I | TESS | 2458442.4185 | 0.0029 | II | TESS |
| 2458958.7196 | 0.0012 | II | TESS | 2458443.3092 | 0.0004 | I | TESS |
| 2458960.1330 | 0.0012 | I | TESS | 2458444.1951 | 0.0006 | II | TESS |
| 2458961.5477 | 0.0014 | II | TESS | 2458445.0783 | 0.0008 | I | TESS |
| 2458962.9591 | 0.0011 | I | TESS | 2458445.9656 | 0.0005 | II | TESS |
| 2458964.3774 | 0.0011 | II | TESS | 2458446.8491 | 0.0010 | I | TESS |
| 2458965.7857 | 0.0008 | I | TESS | 2458448.6207 | 0.0005 | I | TESS |
| 2458967.1992 | 0.0010 | II | TESS | 2458449.5064 | 0.0010 | II | TESS |
| 2458970.0246 | 0.0009 | II | TESS | 2458452.1615 | 0.0009 | I | TESS |
| 2458971.4390 | 0.0011 | I | TESS | 2458453.0460 | 0.0010 | II | TESS |
| 2458972.8509 | 0.0022 | II | TESS | 2458453.9322 | 0.0005 | I | TESS |
| 2458974.2649 | 0.0011 | I | TESS | 2458454.8171 | 0.0011 | II | TESS |
| 2458975.6780 | 0.0017 | II | TESS | 2458455.7002 | 0.0009 | I | TESS |
| 2458977.0923 | 0.0008 | I | TESS | 2458456.5881 | 0.0008 | II | TESS |
| 2458978.5039 | 0.0012 | II | TESS | 2458457.4725 | 0.0008 | I | TESS |
| 2458979.9181 | 0.0006 | I | TESS | 2458458.3557 | 0.0010 | II | TESS |
| 2458981.3324 | 0.0011 | II | TESS | 2458459.2408 | 0.0011 | I | TESS |
| 2460521.5204 | 0.0091 | II | ZC600 | 2458460.1264 | 0.0013 | II | TESS |
| 2460520.4465 | 0.0010 | I | ZC600 | 2458461.0128 | 0.0010 | I | TESS |
| | | | | 2458461.9004 | 0.0008 | II | TESS |
| | | | | 2458462.7821 | 0.0010 | I | TESS |

0.000016  0.000002$^d$. Таким чином, значення періоду може дорівнювати 2.826656  0.000002$^d$, але необхідні подальші дослідження його стабільності.

Для NSV 1884 (рис. 3*в*, 3*г*), схоже, не потрібна корекція періоду в порівнянні з елементами, обчисленими тільки за даними TESS (оскільки точки, що відповідають даним VSX та NSVS, свідчать про протилежні зсуви величини періоду).

## ВИСНОВКИ

Ми вивчили чотири малодосліджені змінні зорі, що мали ймовірну класифікацію типу Алголя. Для двох з них ми знайшли періоди змінності вперше. Зоря NSV 575 виявилась типовою змінною типу Алголя з періодом 0.953578$^d$, натомість NSV 14, скоріш за все, не є затемнюваною, оскільки показує плавну асиметричну хвилю з періодом 13 діб, подібно до пульсуючої змінної. Для NSV 489 та NSV 1884 ми відкоригували відомі періоди та зібрали моменти мінімумів для подальшого аналізу $O - C$. Для трьох затемнюваних змінних ми обчислили пара-метри затемнень. Дані TESS виявились достатньо точними, щоб дозволити нам описати тонкі ефекти відбиття (NSV 575 та NSV 489) та еліптичності зір (NSV 489 and NSV 1884).

**ПОДЯКИ**

Ми вдячні Астрономічній обсерваторії на Колоницькому Сідлі та Обсерваторії і Планетарію у Глоговці за організацію астрономічного табору «Variable-2024» та надання часу для спостережень на їхніх телескопах, та Теодору Прибуллі за організацію лекцій та практичних занять під час цього табору. Ця робота була проведена також в рамках більш широких проєктів (без додаткового фінансування) «Астроінформатика та Віртуальна обсерваторія» [22] та «Міждовготна астрономія» [5]. Ця стаття містить дані, зібрані під час місії TESS [20], а також з оглядів ASAS-SN [16], NSVS [23]. Фінансування місії TESS забезпечує програма NASA Explorer.

**REFERENCES**

1. Andronov I. L. (1994) (Multi-) Frequency variations of stars. Some methods and results. *Odessa Astron. Publ.* 7. 49—54.
2. Andronov I. L. (2003) Multiperiodic versus noise variations: mathematical methods. *Astron. Soc. Pacif. Conf. Ser*. 292. 391.
3. Andronov I. L. (2012) Phenomenological modeling of the light curves of algol-type eclipsing binary stars. *Astrophys.* 55. 536—550. DOI: 10.1007/s10511-012-9259-0.
4. Andronov I. L. (2020) Advanced time series analysis of generally irregularly spaced signals: Beyond the oversimplified methods. Knowledge discovery in big data from astronomy and Earth observation. 191. DOI:10.1016/B978-0-12-819154-5.00022-9.
5. Andronov I. L., Andrych K. D., Antoniuk K. A., et al. (2017) Instabilities in Interacting Binary Stars. *Astron. Soc. Pac. Conf. Ser*. 511. P. 43. DOI 10.48550/arXiv.1702.02011.
6. Andronov I. L., Baklanov A. V. (2004) Algorithm of the artificial comparison star for the CCD photometry. *Astronomical School's Report.* 5. P. 264–272. DOI: 10.18372/2411-6602.05.1264
7. Andronov I. L., Tkachenko M. G., Chinarova L. L. (2017) Comparative Analysis of Phenomenological Approximations for the Light Curves of Eclipsing Binary Stars with Additional Parameters. *Astrophys*. 60, N4. P.57. DOI: 10.1007/s10511-017-9462-0.
8. Andrych K. D., Andronov I. L., Chinarova L. L. (2017) Statistically optimal modeling of flat eclipses and exoplanet transitions. The wall-supported polynomial" (WSP) Algoritms. *Odessa Astron. Publ*. 30. P. 57-62. DOI: 10.18524/1810-4215.2017.30.118521.
9. Andrych K. D., Andronov I. L., Chinarova L. L. (2020) MAVKA: Program of statistically optimal determination of phenomenological parameters of extrema. Parabolic spline algorithm and analysis of variability of the semi-regular star Z UMa. *J. Phys. Stud.* 24, N 1. P. 1902. DOI: 10.30970/jps.24.1902.
10. Boyd D. (2012) VSX page. URL: https://www.aavso.org/vsx_docs/39113/121/Phase%20plot_1.jpg.

11. Drake A. J., Djorgovski S. G., Mahabal A., et al. (2009) First results from the Catalina Real-time Transient Survey. *Astrophys. J.* 696, N 1. P. 870–884. DOI: 10.1088/0004-637X/696/1/870.
12. Gaia Collaboration, Vallenari A., Brown A. G. A., et al. (2023) Gaia Data Release 3: Summary of the content and survey properties. *Astron. and Astrophys*. 674. A1, 22. DOI: 10.1051/0004-6361/202243940
13. Hoffmeister C. (1933) 115 neue Veränderliche. *Astron. Nachr*. 247, N 15. 281—284. DOI: 10.1002/asna.19322471502.
14. Hoffmeister C. (1963) Veränderliche Sterne am Südhimmel. *Veröff. Sternw. Sonneberg*. 6. P. 1—63.
15. International Variable Star Index. (2025) URL: https://www.aavso.org/vsx/
16. Kochanek C. S., Shappee B. J., Stanek K. Z. (2017) The all-sky automated survey for Supernovae (ASAS-SN) light curve server v1.0. *Publs Astron. Soc. Pacif.* 129, 104502. DOI: 10.1088/1538-3873/aa80d9.
17. Marsakova V. I., Andronov I. L. (1996) Local fits of signals with asymptotic branches. *Odessa Astron. Publ*. 9. 127—130.
18. Mikulášek Z. (2015) Phenomenological modelling of eclipsing system light curves. *Astron. and Astrophys*. 584. A8, 13. DOI: 10.1051/0004-6361/201425244.
19. Richter G. A. (1969) Neue Veränderliche. *Mitteil. Veränderliche Sterne*. 5. 69—72.
20. Ricker G. R., Winn J. N., Vanderspek R., et al. (2014) Transiting Exoplanet Survey Satellite (TESS). *Proc. SPIE.* 9143, id. 914320, 15. DOI: 10.1117/12.2063489.
21. Tkachenko M. G., Andronov I. L., Chinarova L. L. (2016) Phenomenological parameters of the prototype eclipsing binaries Algol, Lyrae and W UMa. *J. Phys. Stud.* 20, N 4. 4902. DOI: 10.30970/jps.20.4902.
22. Vavilova I. B., Yatskiv Ya. S., Pakuliak L. K., et al. (2017) UkrVO Astroinformatics software and web-services. IAU Symp. 325. 361—362. DOI: 10.1017/S1743921317001661
23. Woźniak P. R., Vestrand W. T., Akerlof C. W., et al. (2004) Northern Sky Variability Survey: Public data release. *Astron. J.* 127. 2436—2449. DOI: 10.1086/382719.

*V. I. Marsakova*[1,2], *I. L. Andronov*[3], *V. O. Borshchenko*[3], *I. A. Garbazhii-Romanchenko*[1], *A. D. Lashkova*[4], *S. A. Kreminska*[3], *P. A. Dubovsky*[5], *V. V. Dubovskyi*[6], *K.Petrik*[7]

[1]Richelieu Science Lyceum
5 Universytetska St., Odesa, Ukraine, 65026
[2]Main Astronomical Observatory of National Academy of Sciences of Ukraine
27 Akademika Zabolotnoho St., Kyiv, Ukraine, 03143
[3]Odesa National Maritime University
34 Mechnikova str., Odesa, Ukraine, 65029
[4]Pavol Jozef Šafárik University
2 Šrobárova, 04180, Košice, Slovakia
[5]Vihorlat Observatory in Humenne
4 Mierov<, 06601, Humenné, Slovakia
[6]Odesa I. I. Mechnikov National University
2 Vsevoloda Zmienka str., Odesa, Ukraine, 65000
[7]M. R. Stefanik Observatory and Planetarium
41 Sladkovicova, SK-92001 Hlohovec, Slovakia

DETERMINATION OF LIGHT CURVE PARAMETERS OF POORLY STUDIED ECLIPSING VARIABLES USING DATA FROM TESS AND OTHER SKY SURVEYS

A group of poorly studied eclipsing variables (the classification of which is marked as uncertain and/or the period of brightness changes is uncertain) has been studied with the using of the photometric observations of the TESS mission and NSVS, ASAS-SN sky-surveys. We also obtained some observations covering the brightness minima of our variables by our group using the telescopes at Astronomical Observatory on Kolonica Saddle (Slovakia) and Observatory and Planetarium in Hlohovec (Slovakia) during the "Variable-2024" astrocamp. The periods and classification were corrected. For NSV 575 and NSV 014 the periods were found for the first time, but it is doubtful that NSV 014 is an eclipsing variable, because there are no eclipses but the asymmetric wave is present, which indicates that the variable star can be re-classified as a low-amplitude pulsating one. Different methods were used for approximation of the light curves and further calculation of stellar system's parameters such as eclipse depths and durations, values of reflection effect and effect of ellipticity of stars. The initial period was estimated using the periodogram based on the trigonometrical polynomial fit of high order (up to 10). For better approximation of the complete eclipsing phase curve, the "New Algol Variable" (NAV) software was used. The methods of "asymptotic parabolas" and "wall-supported asymptotic parabolas" were used for calculation of moments of eclipses, which use only near-eclipse part of the observations instead of a complete curve. These methods were implemented in the software MAVKA among a larger set of features. For the variables NSV 489 and NSV 1884, our moments of eclipses and the ones found in the literature, were used for the $O - C$ curves. For NSV 489, the period was adjusted taking into account the slope of the $O - C$ diagram.

***Keywords:*** astroinformatics, eclipsing variables, binary stars, data analysis, light curve analysis, TESS, NSVS, ASAS-SN